\documentclass[aps,pre,onecolumn,nofootinbib,11pt,tightenlines]{revtex4-1}
\usepackage{amsmath,amssymb,amstext,graphicx,color,xfrac,braket,accents,tabulary,booktabs,xcolor,subcaption}
\begin{document}
\baselineskip=14pt
\hfill CALT-TH-2018-010
\hfill

\vspace{0.5cm}
\thispagestyle{empty}

\title{Towards Space from Hilbert Space: Finding Lattice Structure in Finite-Dimensional Quantum Systems }
\author{Jason Pollack}
\email{jpollack@phas.ubc.ca}
\affiliation{Department of Physics and Astronomy, University of British Columbia, Vancouver, BC, V6T 1Z1, Canada}
\author{Ashmeet Singh}
\email{ashmeet@caltech.edu}
\affiliation{Walter Burke Institute for Theoretical Physics, California Institute of Technology, Pasadena, CA 91125}

\newcommand{\be}{\begin{equation}}
\newcommand{\ee}{\end{equation}}
\newcommand{\Ref}[1]{Ref.~\cite{#1}}
\newcommand{\Fig}[1]{Fig.~\ref{#1}}
\newcommand{\Eq}[1]{Eq.~\eqref{#1}}
\newcommand{\Eqs}[2]{Eqs.~\eqref{#1} and \eqref{#2}}
\newcommand{\Sec}[1]{Sec.~\ref{#1}}
\newcommand{\Secs}[2]{Secs.~\ref{#1} and \ref{#2}}
\newcommand{\App}[1]{App.~\ref{#1}}

\newcommand{\Hil}{{\cal H}}
\newcommand{\Op}{{\hat{\mathcal O}}}
\newcommand{\ham}{{\hat H}}
\newcommand{\Hext}{{\cal H}_{\rm ext}}
\newcommand{\Hint}{{\cal H}_{\rm int}}
\newcommand{\N}{{\mathbb N}}
\newcommand{\Dim}{\textrm{dim\,}}
\newcommand{\Tr}{\textrm{Tr\,}}
\newcommand{\pr}{\textrm{Pr}}
\newcommand{\hsST}{\Hil_{\mathrm{spacetime}}}
\newcommand{\opq}{\hat{q}}
\newcommand{\opp}{\hat{p}}
\newcommand{\hsL}{\Hil_{\mathrm{left}}}
\newcommand{\hsR}{\Hil_{\mathrm{right}}}
\newcommand{\eye}{\mathbb{I}}
\newcommand{\tunnel}{\hat{H}_{\mathrm{tunnel}(\sigma)}}
\newcommand{\spsi}{\ket{\psi}}
\newcommand{\sphi}{\ket{\phi}}
\newcommand{\oppi}{\hat{\pi}}


\newcommand{\hst}{\widetilde{\mathcal{H}}} 
\newcommand{\iso}{\dot{=}}
\newcommand{\tdec}{{\{\theta\}}}
\newcommand{\pdec}{{\{\phi\}}}

\newcommand{\psimu}{\ket{\psi^{(\mu)}}}
\newcommand{\Cmu}{\left[ C^{(\mu)} \right]}
\newcommand{\Y}{\left[ C \right]}
\newcommand{\Ymean}{\left[ \bar{C} \right]}
\newcommand{\deltaY}{\left[ \Delta C \right]}
\newcommand{\OD}{\left[ O_{D} \right]}
\newcommand{\phiM}{\left[ \Phi \right]}

\newcommand{\hs}{\mathcal{H}} 
\newcommand{\A}{\hat{A}}
\newcommand{\B}{\hat{B}}
\newcommand{\U}{\hat{U}}
\newcommand{\V}{\hat{V}}
\newcommand{\opphi}{\hat{\phi}}

\newcommand{\D}{\hat{D}}
\newcommand{\pos}{\hat{Q}}
\newcommand{\mom}{\hat{P}}

\newcommand{\Tu}{\hat{T}_{u}}
\newcommand{\Tv}{\hat{T}_{v}}
\newcommand{\subA}{\mathcal{A}}
\newcommand{\subB}{\mathcal{B}}
\newcommand{\oprho}{\hat{\rho}}
\newcommand{\intham}{\hat{H}_{\rm{int}}}
\newcommand{\selfham}{\hat{H}_{\rm{self}}}
\newcommand{\Asig}{A_{\sigma}}
\newcommand{\Bsig}{B_{\sigma}}
\newcommand{\M}{\hat{M}}

\def\({\left(}
\def\){\right)}
\def\[{\left[}
\def\]{\right]}
\def\llangle{\left\langle}
\def\rrangle{\right\rangle}

\newcommand{\eins}{\mbox{$1 \hspace{-1.0mm} {\bf l}$}}

\newcommand{\draftnote}[1]{\textbf{\color{blue}[#1]}}

\begin{abstract}
Field theories place one or more degrees of freedom at every point in space.
Hilbert spaces describing quantum field theories, or their finite-dimensional discretizations on lattices, therefore have large amounts of structure: they are isomorphic to the tensor product of a smaller Hilbert space for each lattice site or point in space.
Local field theories respecting this structure have interactions which preferentially couple nearby points.
The emergence of classicality through decoherence relies on this framework of tensor-product decomposition and local interactions.
We explore the emergence of such lattice structure from Hilbert-space considerations alone.
We point out that the vast majority of finite-dimensional Hilbert spaces cannot be isomorphic to the tensor product of Hilbert-space subfactors that describes a lattice theory.
A generic Hilbert space can only be split into a direct sum corresponding to a basis of state vectors spanning the Hilbert space; we consider setups in which the direct sum is naturally decomposed into two pieces.
We define a notion of direct-sum locality which characterizes states and decompositions compatible with Hamiltonian time evolution.
We illustrate these notions for a toy model that is the finite-dimensional discretization of the quantum-mechanical double-well potential.
We discuss their relevance in cosmology and field theory, especially for theories which describe a landscape of vacua with different spacetime geometries.

\end{abstract}

\maketitle

\tableofcontents

\section{Introduction}

Mathematically, the basic objects of quantum mechanics are state vectors in an abstract Hilbert space.
Yet the real world is well-described by one such state in one such space. 
It is natural to ask what additional features distinguish our state and Hilbert space from generic ones.
We know, for example, some details of the field content of our universe: it contains (at minimum) the fields in the Standard Model, a spin-two graviton, and potentially additional fields such as dark matter, an inflaton, etc.
In particular, the quantum-mechanical theory which describes our universe has a description (in the semiclassical limit) as a theory of \emph{fields}: that is, it has degrees of freedom which live at each point of some background spatial manifold (which in turn is a spatial slice of a four-dimensional spacetime geometry which solves Einstein's equations).
Furthermore, to a very good approximation the universe appears classical: we typically observe objects with definite values of classical variables (such as position and momentum) rather than in superpositions, and the time evolution of (expectation values of) these quantities obeys classical equations of motion.
When considered as a point in a classical Hamiltonian phase space, it is also apparent that the current state of the universe is special: it is a low-entropy state far from equilibrium, with nontrivial evolution that exhibits an arrow of time.

Understanding the origin of all of these features is a vast research program.
In this paper we focus on one feature: the fact that the time evolution of the state vector of the universe can be described as the time evolution of field-theoretic degrees of freedom living on a background space of definite dimension (and geometry).
We are motivated to investigate this feature in particular because it seems to be a prerequisite for applying our most successful models of the emergence of classicality.
The decoherence program \cite{Zeh:1970fop,Zurek:1981xq,Griffiths:1984rx,Joos:1984uk,Schlosshauer:2003zy} explains how the unitary evolution of a single quantum-mechanical state is naturally viewed as a process involving the creation (via entropy production) of distinct classical branches which evolve independently without interference. 
The set of branches is selected by the Hamiltonian governing time evolution: when Hilbert space is decomposed into a preferred choice of subsystems \cite{Piazza:2005wm,mereology}, the branches are the states which remain robust to the influence of the interactions between subsystems, i.e.\ in which the state of a given subsystem is preserved by interactions with the environment.
This story relies crucially on the ability to decompose the Hilbert space into many interacting subsystems---or, equivalently, to identify local degrees of freedom \cite{Cotler:2017abq}.
Once these local degrees of freedom are identified, it seems plausible that space itself can be built up from considering the interactions between subsystems (c.f.\ \cite{2012NJPh...14h3010J,Cao:2016mst,er-eprmvr,Carroll:2018rhc} and references therein), although this process is still incompletely understood.
Or, more directly, the degrees of freedom can be organized into a spatial lattice or spin chain.

\begin{samepage}
The goal of this paper is to provide answers to two questions:
\begin{itemize}
\item When does a quantum-mechanical theory describe spatial degrees of freedom?
\item When we know a theory \emph{does} descibe spatial degrees of freedom, to what extent can we identify them from purely quantum-mechanical data?
\end{itemize}
\end{samepage}

In investigating these questions we largely restrict ourselves to finite-dimensional Hilbert spaces.
This is partly for convenience: understanding how such spaces can be decomposed is much more mathematically tractable (with no need, for example, to consider type III von Neumann algebras).
Nevertheless, a number of arguments associated with complementarity and black hole entropy \cite{Bekenstein:1980jp,Susskind:1993if,Srednicki:1993im} suggest that the set of degrees of freedom accessible to any observer in a local region of space is actually finite \cite{Bao:2017rnv}. These arguments are sharpest in an asymptotically de~Sitter spacetime which is dominated by vacuum energy, where a horizon-sized patch of spacetime is a maximum-entropy thermal state with a finite entropy and a corresponding finite number of degrees of freedom \cite{Banks2000,Fischler2000}.

Given this restriction, we can answer the first question by checking when a finite-dimensional quantum-mechanical theory can describe a lattice theory.
A simple number-theoretic argument, which we give in Section \ref{sec:non-genericity} below, gives a surprising answer to this question: almost never!
That is, for almost all choices of finite positive integer $N$, independent of the Hamiltonian, there is \emph{no} Hilbert space of dimension $N$ which can describe a lattice theory with spatial dimension $\ll N$.
We are therefore led to slightly generalize our setup, to include Hilbert spaces which can be decomposed into pieces which each describe spatial lattices.
As a toy model, we consider the finite-dimensional analog of the double-well potential.
For a large enough barrier, low-lying states should decompose into a piece in the left well and a piece in the right well.
We use the tools of generalized Clifford algebras (GCAs) (for a review, see \cite{Jagannathan:2010sb} and references therein) to formalize this intuition.
The lessons from this simple example should be applicable to more general examples of cosmological relevance, such as landscape potentials in which each minimum describes a different metastable vacuum solution.

The remainder of this paper is organized as follows.
In Section \ref{sec:non-genericity} we give a simple number-theoretic argument that almost all finite-dimensional Hilbert spaces are unable to describe lattice theories.
In Section \ref{sec:direct-sum} we therefore move on to describe Hilbert spaces which are a direct sum of lattice theories.
In Section \ref{sec:locality} we build on this description to give a definition of \emph{direct-sum locality}, which measures when a particular decomposition of a Hilbert space divides it into pieces which remain separate under the action of the Hamiltonian. 
In Section \ref{sec:double-well} we apply these definitions to a worked example: the double-well potential.
We show how we can use the various measures of locality to identify a natural decomposition of the Hilbert space which successfully describes a spatial lattice theory.
In Section \ref{sec:field-theory} we argue that the strategy developed for the double-well potential should be of more general applicability to (finite-dimensional truncations of) field theory.
Finally, we conclude in Section \ref{sec:conclusion}.
\section{The Non-Genericity of Lattice Hilbert Spaces} \label{sec:non-genericity}

Consider a finite-dimensional quantum-mechanical theory that lives on a spatial lattice, i.e.\ whose Hilbert space is isomorphic to a tensor product of smaller Hilbert spaces (we have used $\simeq$ to denote Hilbert space isomorphism),
\be
\Hil_{\rm lattice} \simeq \Hil_{\rm site}^{\otimes N_{\rm sites}}\label{eq:H_lattice}.
\ee
If the lattice is embedded in a multidimensional space, we can further write
\be
N_{\rm sites} = \prod_{i=1}^{N_{\rm dim}} N_{\rm sites}^{(i)}.
\ee

Now consider the constraints that the factorization relation \eqref{eq:H_lattice} places on the dimensionality of $\Hil_{\rm lattice}$.
We have seen that the dimension of the Hilbert spaces we are considering takes the form
\be
\left|\Hil_{\rm lattice}\right| = \left|\Hil_{\rm site}\right|^{N_{\rm sites}}.
\ee
So, just as the Hilbert space has $N_{\rm sites}$ subfactors, its dimension has $\sim N_{\rm sites}$ prime factors (where the $\sim$ covers the fact that $|\Hil_{\rm site}|$ might itself have multiple prime factors).
That is, 
\be
\textrm{\# of prime factors of }\left|\Hil_{\rm lattice}\right|
\equiv \Omega\( \left|\Hil_{\rm lattice}\right| \) \sim \ln \left|\Hil_{\rm lattice}\right|.
\ee

The function $\Omega(n)$ counts the number of prime factors (including multiplicity) of the natural number $n$.
It is closely related to $\omega(n)$, the number of distinct prime factors of $n$.
Famously, the Hardy-Ramanujan theorem \cite{zbMATH02610444} says that asymptotically
\be
\omega\(n\) \sim \ln \ln n, \mathrm{var}\(\omega\(n\)\) \sim \ln \ln n.
\ee
The total number of prime factors $\Omega(n)$ can be shown to have a similar asymptotic expansion (e.g.\  \cite{JLMS:JLMS0308}):
\be
\Omega\(n\) \sim \ln \ln n, \mathrm{var}\(\Omega\(n\)\) \sim \ln \ln n.
\ee

So, as the size of a Hilbert space gets larger, it becomes vanishingly rare for the Hilbert space to have a dimension of the right size for it to describe a lattice theory.
\begin{figure}[t]
  \begin{center}
    \includegraphics[width=0.9\textwidth]{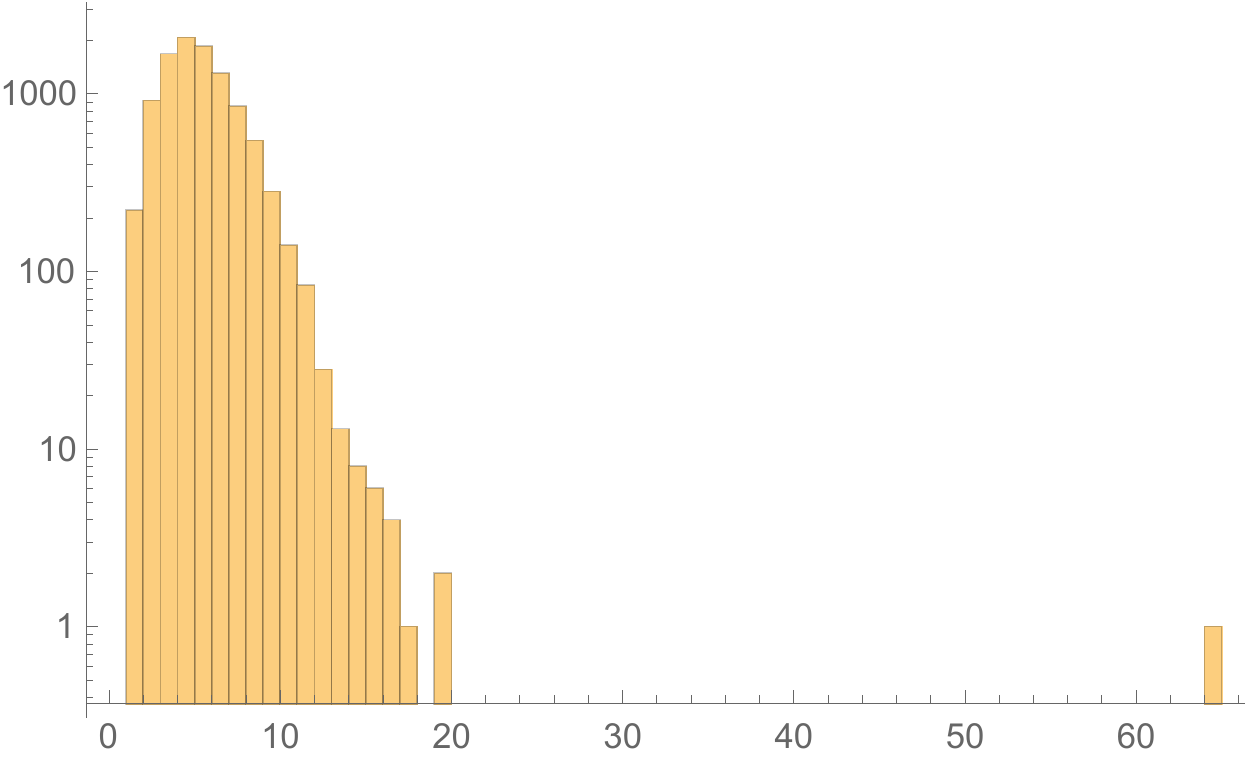}
  \end{center}
 \caption{Histogram of $\Omega(n)$ for $2^{64}-5000 < n < 2^{64} + 5000$.
  The mean is 4.85, the standard deviation is 2.21.\label{fig:hist1}}
\end{figure}

\begin{figure*}[t!]
    \centering
    \begin{subfigure}[b]{0.5\textwidth}
        \centering
        \includegraphics[height=2.1in]{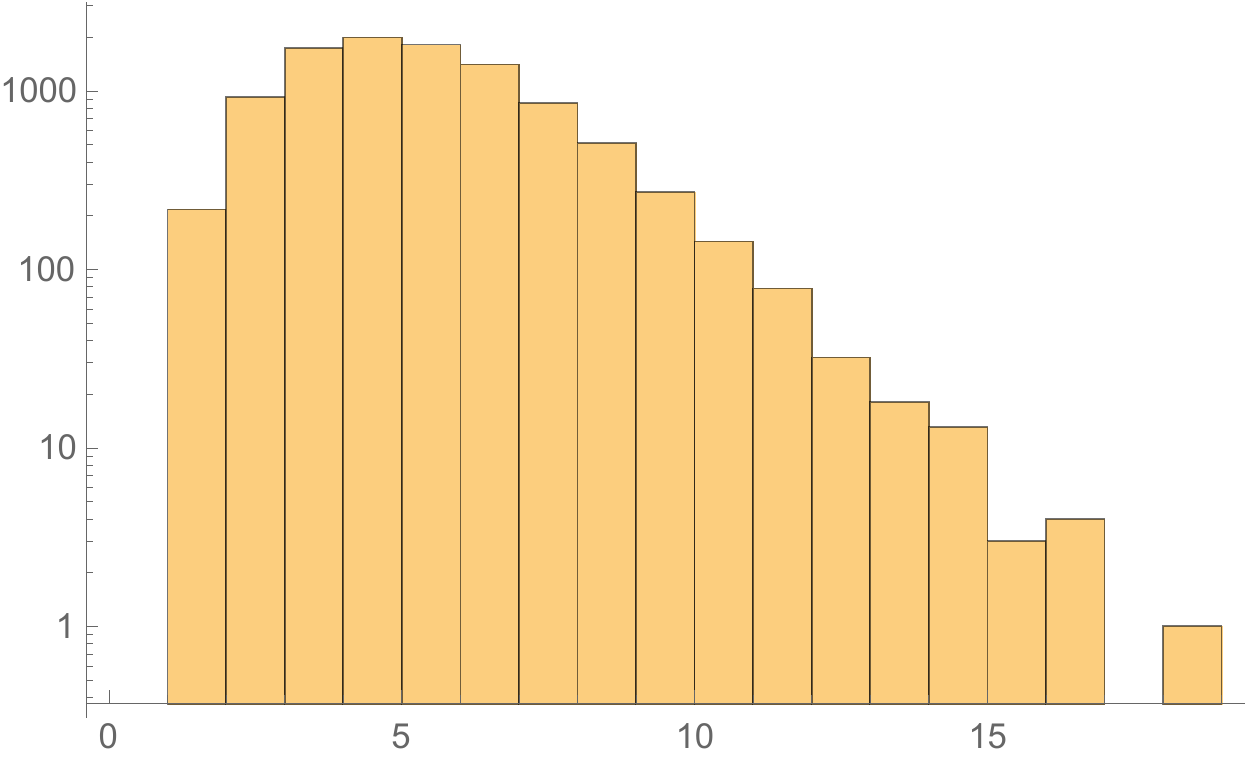}
        \caption{$2^{64}-15000 < n < 2^{64} - 5000$.}
    \end{subfigure}%
    ~ 
    \begin{subfigure}[b]{0.5\textwidth}
        \centering
        \includegraphics[height=2.1in]{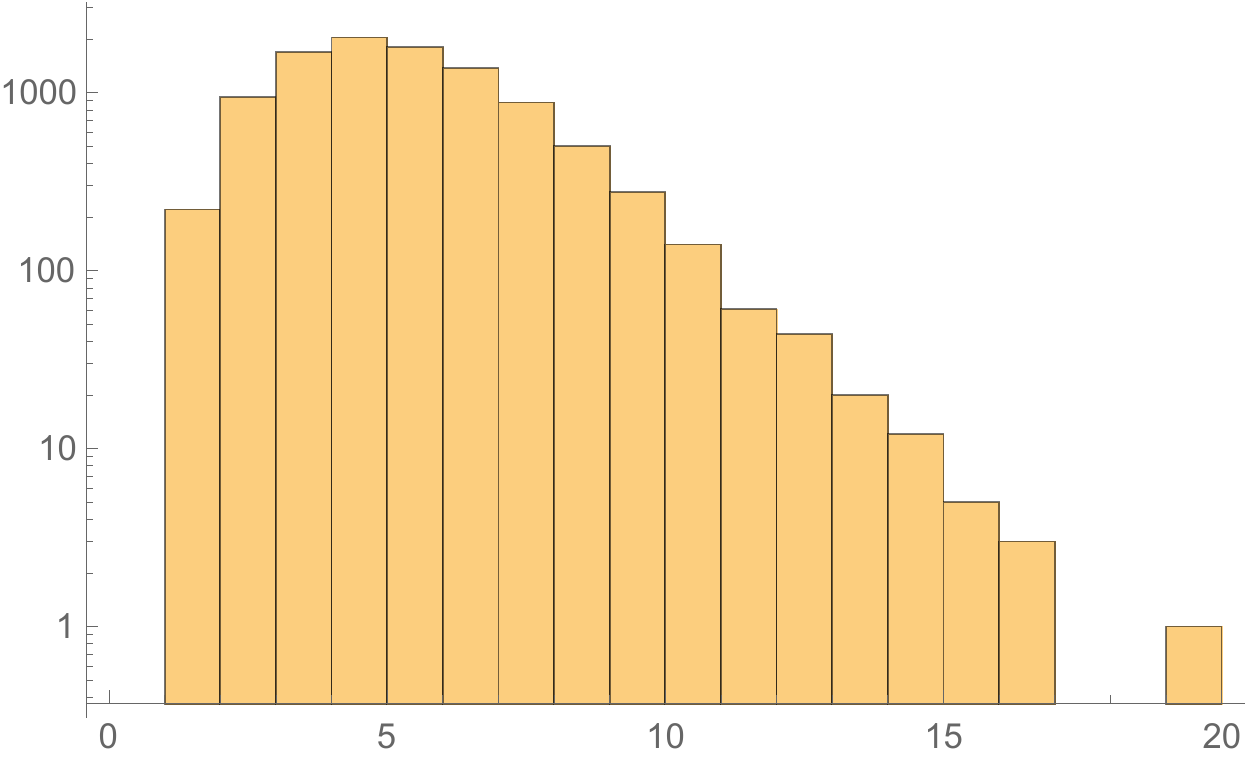}
        \caption{$2^{64}+5000 < n < 2^{64} + 15000$.}
    \end{subfigure}
    \caption{Histogram of $\Omega(n)$ for different ranges of $n$. The mean is 4.84, the standard deviation is 2.13 in both cases.\label{fig:hist2}}
\end{figure*}

To gain some intuition for this phenomenon, consider Hilbert spaces around the same size as that of a $4\times4\times4$ lattice of qubits, 
\be
\left|\Hil_{\rm lattice}\right|\approx 2^{64}\approx 1.8 \times 10^{19}.
\ee
We have 
\be
\ln \left|\Hil_{\rm lattice}\right| \approx 64 \ln 2 \approx 44, \ln\ln \left|\Hil_{\rm lattice}\right| \approx 3.8.
\ee
As Figures \ref{fig:hist1} and \ref{fig:hist2} show, when we histogram the integers around $2^{64}$ we indeed find that typical integers $n$ in this range have $\Omega(n)\sim\ln \ln n$.
In particular, the mean number of factors is $4.8$ and the standard deviation around the mean is $2.1-2.2$.
$2^{64}$ itself is then an extreme---$30$ sigma!---outlier.
\section{Direct-Sum Hilbert Spaces} \label{sec:direct-sum}

In the previous section we argued that large Hilbert spaces---like the one which describes our own universe, which must be at least as large as $\exp(S_{dS})\sim\exp(10^{122})$ to describe our Hubble volume---are vanishingly unlikely to decompose in the manner necessary to describe a lattice quantum field theory.
On the other hand, we can always identify subspaces of a large Hilbert space---for example, those with dimension equal to the largest power of two smaller than the dimensionality of the Hilbert space---which might themselves be decomposed into a product over lattice sites:
\be
\Hil=\Hil_{\mathrm{lattice}} \oplus \Hil_\textrm{remainder}, \:
\left|\Hil_{\mathrm{lattice}}\right|=2^{\lfloor \log_2 n \rfloor}\equiv n_{\mathrm{lattice}} 
\Longrightarrow \Omega(n_{\mathrm{lattice}})\sim \ln n_{\mathrm{lattice}} \sim \ln \left|\Hil\right|.
\ee
Could the Hilbert space of our universe be of this form? 
In such a situation a generic state in $\Hil$ would be a superposition of a state in the lattice subspace and a state in the (typically non-geometric) remainder space.
Put another way, an initial ``geometric'' state in the lattice Hilbert space is not constrained to remain within it under the action of the Hamiltonian: part of it can ``leak out'' into the remainder of the Hilbert space.
This is not a familiar situation in standard quantum field theory, where the use of a unitary S-matrix is predicated on both initial and final asymptotic states being the vacuum of a field theory on a fixed background.
However, we have used language meant to suggest situations where this does occur:  barrier decay in quantum mechanics or, in quantum field theory in curved space, the decay of metastable vacua by bubble nucleation.

In the latter case, one typically considers states localized around particular (meta)stable vacua, which are each given geometric interpretations, but a generic state describes a superposition of field configurations of different background geometries. 
That is, states in such a theory with multiple vacua are necessarily not states in a single field theory, but superpositions of states in different field theories, and thus a finite-dimensional version of such a theory does not have a Hilbert space with the tensor-product structure of a single lattice theory but is instead a sum of such tensor-product spaces.
We discuss the field-theoretic interpretations of our results further in Section \ref{sec:field-theory} below.
In the remainder of this section, we develop a formalism for the simplest such system: Hilbert spaces which divide into two pieces, each of which describes a lattice.

Suppose we have a Hilbert space $\Hil$ with a finite dimension $\mathrm{dim}\:\Hil = N < \infty$. Given an operator $\Op \in \mathcal{L}(\Hil)$, we can write the eigenstates of the operator
\be
\Op \ket{o_i} = O_i \ket{o_i}
\ee
(dealing with degeneracies as necessary so that the $\{ \ket{o_{i}}   \: , \: i = 1,2,\ldots,N \}$ are an orthonormal basis for $\Hil$) and decompose a generic state $\ket{\Psi}\in \Hil$ as
\be
\ket{\Psi}=\sum_i c_i^{(\Op)} \ket{o_i}.
\ee
We can thus also decompose the Hilbert space as a direct sum of one-dimensional subspaces,
\be
\Hil = \bigoplus_i \Hil_i^{(\Op)},
\ee
with $\Hil_i^{(\Op)}\simeq \mathbb{C}$ the one-dimensional Hilbert space consisting of scalar multiples of $\ket{o_i}$.
Let us define a choice of \emph{scrambling} of this direct-sum decomposition by choosing a permutation $\sigma$ of our set of ordered eigenstates $\(1,2,3,\ldots,N\)$ followed by a division into two mutually exclusive, and exhaustive sets $A_{\sigma}$ and $B_{\sigma}$, of cardinality $|A_{\sigma}| = m$ and $|B_{\sigma}| = N -m$,
 \be
 \sigma\(1,\ldots,N\)=\(\sigma_1,\ldots,\sigma_N\)=\(\sigma_1,\ldots,\sigma_m\)\cup\(\sigma_{m+1},\ldots,\sigma_N\)\equiv A_\sigma \cup B_\sigma \: ,
 \ee
where, of course, viewed as unordered sets we have $A_{\sigma} \cup B_{\sigma} = \{1,2,3,\ldots,N \}$ and $A_{\sigma} \cap B_{\sigma} = \varnothing$. We will denote the canonical, ordered set $\(1,2,3,\ldots,N \)$ as the one corresponding to $\sigma = {\mathrm{id}}$ (i.e. the identity permutation). This allows us to write our Hilbert space $\Hil$ as a direct sum of two Hilbert spaces of dimension $m$ and $N-m$, respectively:
\begin{equation}
\label{Hilbert_direct_sum_def}
\Hil \simeq \Hil^{\(\Op\)}_{A_\sigma} \oplus \Hil^{\(\Op\)}_{B_\sigma} \: ,
\end{equation} 
where
\be
\Hil^{\(\Op\)}_{A_\sigma} \equiv \bigoplus_{j \in A_\sigma} \Hil^{(\Op)}_{j} \: \: \:  \mathrm{and} \: \: \:   \Hil^{\(\Op\)}_{B_\sigma} \equiv \bigoplus_{j' \in B_\sigma} \Hil^{(\Op)}_{j'} .
\ee
Let us denote this choice of direct-sum decomposition as $D_{\oplus} \equiv \{\Op,\sigma,m \}$ which consists of a choice of the operator $\Op$, the permutation $\sigma$ and subspace size $m$ as defined above.

Our formalism thus far has been totally generic: it describes all possible partitions of a Hilbert space of dimension $N$ into two parts. 
We would like, however, to find the particular partitions (if there are any) which reflect genuine features of the theory.
In particular, as we discussed above, useful partitions should be approximately preserved under time evolution, so that geometric states localized in one of the Hilbert spaces only gradually leak into the other one.
We say that partitions where this is the case exhibit \emph{direct-sum locality}.
To diagnose it we will need measures which depend not only on $N$ and $\Op$ but also on the Hamiltonian $\ham$.

First consider the special case $\Op = \ham$.
We can decompose any state $\ket{\Psi}$ in terms of energy eigenstates $\ket{e_i}$, where each $\Hil_i^{(\ham)}$ is (isomorphic to) the set of vectors in $\mathbb C^N$ proportional to $\ket{e_i}$.
A natural division of the energy eigenstates is the states below/above the energy $E_m$ of the $m$-th energy eigenstate (with the energy eigenstates arranged in an ascending order with $\ket{e_{1}}$ being the ground state): 
\be
A_{0} = \{1,2,\ldots,m\}
\ee
\be
\Hil^{(\ham_{\!A_{0}})} = \bigoplus_{j\in A_{0}} \Hil^{(\ham)}_{j}
\ee
This decomposition is (unsurprisingly) trivial with respect to the Hamiltonian: time evolution only evolves states within the subspaces, and there is no interaction between $\Hil^{(\ham_{\!A_{0}})}$ and $\Hil^{(\ham_{\!B_{0}})}$.

Instead, we should consider the action of $\ham$ on a Hilbert space divided generically as $\Hil = \Hil^{(\Op)}_{A_{\sigma}} \oplus \Hil^{(\Op)}_{B_{\sigma}}$. 
To belabor the point, we can write
\be
\ket{o_i}=\sum_j \braket{e_j|o_i}\ket{e_j}
\ee
and
\be
\ket{e_j}=\sum_k \braket{o_k|e_j}\ket{o_k}
\ee
so
\begin{multline}
e^{-i \ham t}\ket{o_i} = \sum_j e^{-i E_j t} \braket{e_j|o_i}\ket{e_j} \\
= \sum_{j,k} e^{-i E_j t} \braket{e_j|o_i}\braket{o_k|e_j}\ket{o_k}
= \sum_k \(\sum_j e^{-i E_j t} \braket{e_j|o_i}\braket{o_k|e_j}\) \ket{o_k},
\end{multline}
i.e. time evolution evolves an eigenstate of $\Op$ into a superposition of eigenstates.
In particular, for generic $\Op$ the time evolution of $\ket{o_i}$ will have support on both $\Hil^{(\Op)}_{A_{\sigma}}$ and $\Hil^{(\Op)}_{B_{\sigma}}$.

Thus, the Hamiltonian $\ham$ for the system under this direct-sum decomposition $D_{\oplus}\left(\Op,\sigma,m\right)$ of Eq. (\ref{Hilbert_direct_sum_def}) can be decomposed into a term $\ham_{\!A_{\sigma}}$ which acts non-trivially only on the part of the state supported on $\Hil^{(\Op)}_{A_{\sigma}}$, a term $\ham_{\!B_{\sigma}}$ acting only on part of states supported on $\Hil^{(\Op)}_{B_{\sigma}}$ and finally a \emph{tunneling} term $\tunnel$ which swaps support between $\Hil^{(\Op)}_{A_{\sigma}}$ and $\Hil^{(\Op)}_{B_{\sigma}}$ (we have suppressed the superscript $(\Op)$ on the terms in the Hamiltonian to avoid clutter in our notation),
\begin{equation}
\label{Ham_direct_decomposition}
\ham = \ham_{\!A_{\sigma}} + \ham_{\!B_{\sigma}} + \tunnel \: .
\end{equation}
One could work in the eigenbasis of $\Op$ to express the Hamiltonian $\ham$ as a matrix and under the scrambling permutation $\{A_{\sigma},B_{\sigma}\}$, in which case the terms $\ham_{A_{\sigma}}$ and $\ham_{B_{\sigma}}$ would represent diagonal blocks while $\tunnel$ would be the off-diagonal piece. In the next section we seek measures of direct-sum locality which depend on this decomposition of the Hamiltonian.

\section{Direct-Sum Locality and Robustness}\label{sec:locality}

In the previous sections we established the rarity of lattice structures in a generic Hilbert space and motivated the use of direct-sum constructions as tools for finding lattice-like factorizations where locality can be made manifest. As discussed in Section \ref{sec:direct-sum}, a finite-dimensional Hilbert space $\Hil$ can be decomposed into a direct sum of two subspaces labeled by $D_{\oplus}(\Op,\sigma,m)$, which is specified by a choice of the operator $\Op$ whose eigenstates are used to define the direct-sum decomposition and a partition of these eigenstates into two sets.  

In this section, we tackle the problem of finding a suitable measure to quantify the \emph{direct-sum locality} of states in the context of a direct-sum decomposition. Locality in such a context means that states which begin localized in one subspace in the decomposition will remain localized under time evolution by the Hamiltonian and not spread substantially into the other direct sum subspace(s). Hence such states evolve mostly unitarily within that subspace with little or no tunneling into other direct-sum subspaces. We emphasize that direct-sum locality is a highly non-generic property, exhibited only by a subset of states in Hilbert space in a particular choice of direct-sum decomposition. 

To make the notion of direct-sum locality concrete, we need to specify what we mean by ``localized in a subspace.'' Consider an arbitrary state $\sphi \in \Hil$, which in general, has non-trivial support on the full Hilbert space. 
We would like to define a super-operator $\pr^{(\Op)}_{\Asig}$ which takes $\sphi$ and returns a state $\sphi_{\Asig}$ living in $\Hil^{(\Op)}_{\Asig}$ which corresponds to the support of $\sphi$ on $\Hil^{(\Op)}_{\Asig}$ (and a similar super-operator $\pr^{(\Op)}_{\Bsig}$). 
The natural tool to use is the projection operator $\hat{P}_{\Asig}$ onto $\Hil^{(\Op)}_{\Asig}\subset\Hil$:
\be
\hat{P}^{(\Op)}_{\Asig} \equiv \sum_{j\in \Asig} \ket{o_{j}}\bra{o_{j}} \: ,
\ee
where as usual for a projector $\(\hat{P}^{(\Op)}_{\Asig}\)^2 = \hat{P}^{(\Op)}_{\Asig}$.
Now $\hat{P}^{(\Op)}_{\Asig}\sphi$ is not a state vector, because it need not have unit norm:
\be
0\le\braket{\phi |\hat{P}^{(\Op)}_{\Asig}|\phi}\le1.
\ee 

When the norm is nonzero, we can recover a normalized state by dividing by the norm.
When the norm is zero, however, there is no unambiguous way to do this. 
This is, in fact, desirable: the norm is zero only when a state $\sphi$ in fact has no support on $\Hil^{(\Op)}_{\Asig}$.
What this means is that our super-operator $\pr^{(\Op)}_{\Asig}$ does not map strictly from states in $\Hil$ onto states in $\Hil^{(\Op)}_{\Asig}$, but onto either states or the null element\footnote{Recall that because Hilbert spaces are vector spaces they have a null element ${\bf 0_\Hil}\in\Hil$, with $||{\bf 0}_\Hil||=0$.
Because state vectors are (equivalence classes of) vectors in the Hilbert space with unit norm, ${\bf 0}_\Hil$ is not itself a physical state, but it is nonetheless an element of the Hilbert space.} ${\bf 0}_{\Hil^{(\Op)}_{\Asig}}\in\Hil_{\Asig}$.
Hence the action of $\pr^{(\Op)}_{\Asig}$ is defined\footnote{To avoid clutter we have neglected a superscript $(\Op)$ on our projected states, e.g.\ writing $\sphi_{\Asig}$ rather than $\sphi^{(\Op)}_{\Asig}$, but it should be understood that any projected state (in any direct-sum subspace) is dependent on the choice of $\Op$.} as follows:
\begin{equation}
\pr^{(\Op)}_{\Asig} \: : \: \Hil \to \Hil^{(\Op)}_{\Asig}\subset \Hil \: \: , \: \: \sphi \mapsto \sphi_{\Asig} \: ,
\end{equation}
with
\begin{equation}
\label{projection1}
\sphi_{\Asig}  = \begin{cases}
\frac{\hat{P}^{(\Op)}_{\Asig} \sphi}{\braket{\phi |\hat{P}^{(\Op)}_{\Asig}|\phi}}, 
& \braket{\phi |\hat{P}^{(\Op)}_{\Asig}|\phi}>0 \\
{\bf 0}_{\Hil^{(\Op)}_{\Asig}},
& \braket{\phi |\hat{P}^{(\Op)}_{\Asig}|\phi} = 0.
\end{cases}
\end{equation}

We can now proceed to quantify the spread of an arbitrary state $\spsi$ in a given direct-sum decomposition $D_{\oplus}(\Op,\sigma,m)$ by projecting it onto $\Hil^{(\Op)}_{\Asig}$ using $\pr^{(\Op)}_{\Asig}$ and checking to what extent time evolution, given by the action of the Hamiltonian $\ham$ (\ref{Ham_direct_decomposition}), evolves the projected state to have non-trivial support on $\Hil^{(\Op)}_{\Bsig}$. For any state $\spsi$, let us take our initial state $\ket{\psi(0)}$ to be the projection of $\spsi$ on $\Hil^{(\Op)}_{\Asig}$, $\ket{\psi(0)}  \equiv \spsi_{\Asig}$, using Eq. (\ref{projection1}).

For concreteness, we will look at small time evolution of this state. This is physically justified since we expect that in arbitrary choices of direct-sum decompositions, generic states $\spsi$ projected down to $\Hil^{(\Op)}_{\Asig}$ will spread over the entire Hilbert space on very short time scales, representing their non-locality and lack of robustness in a direct-sum sense, whereas robust states (whose properties we will discuss below) would stay localized in the subspace they begin with. The time-evolved state, explicitly written to $\mathcal{O}(t^{2})$, is
\begin{equation}
\label{psi_t_1}
\ket{\psi(t)} = \exp{\left( - i \ham t \right)} \ket{\psi(0)} = \left( \eye - i t \ham -\frac{t^2}{2} \ham^{2} + \mathcal{O}(t^{3}) \right) \frac{\hat{P}^{(\Op)}_{\Asig} \spsi}{\braket{\psi|\hat{P}^{(\Op)}_{\Asig}|\psi}}  \: .
\end{equation}
Substituting the form (\ref{Ham_direct_decomposition}) of the Hamiltonian yields
\begin{multline}\label{eq:psi_t}
\ket{\psi(t)} = \frac{\hat{P}^{(\Op)}_{\Asig} \spsi}{\braket{\psi|\hat{P}^{(\Op)}_{\Asig}|\psi}}  - \frac{i t}{\braket{\psi|\hat{P}^{(\Op)}_{\Asig}|\psi}} \left(\ham_{\Asig} + \tunnel \right) \hat{P}^{(\Op)}_{\Asig} \spsi \\ 
 - \frac{t^2}{2 \braket{\psi|\hat{P}^{(\Op)}_{\Asig}|\psi}} \left(\ham_{\Asig}^{2} + \tunnel^{2}  + \ham_{\Bsig}\tunnel + \tunnel \ham_{\Asig} \right) \hat{P}^{(\Op)}_{\Asig} \spsi  + \mathcal{O}(t^{3}),
\end{multline}
where we have simplified the expression using the orthogonality properties of the projected state,
\begin{equation}
\ham_{\Bsig} \hat{P}^{(\Op)}_{\Asig} \:  = \:  \ham_{\Bsig} \ham_{\Asig} \hat{P}^{(\Op)}_{\Asig} \: =  \: \ham_{\Asig} \tunnel \hat{P}^{(\Op)}_{\Asig} \:  = \:  0 \: .
\end{equation}
Recall that the tunneling term $\tunnel$ swaps support of states localized in either subspace, such that its action on states completely localized in $\Hil^{(\Op)}_{\Asig}$ will transform them to states with support only in $\Hil^{(\Op)}_{\Bsig}$ and \emph{vice versa}.

Eq.\ (\ref{eq:psi_t}) makes clear that that the time-evolved state $\ket{\psi(t)}$ has support over the full Hilbert space $\Hil$, as expected due to the presence of the tunneling term $\tunnel$, even though the initial state $\ket{\psi(0)}$ was constructed to be localized only in $\Hil^{(\Op)}_{\Asig}$. We would like to quantify how much support $\ket{\psi(t)}$ has in $\Hil^{(\Op)}_{\Bsig}$. This can be achieved by projecting $\ket{\psi(t)}$ to $\Hil^{(\Op)}_{\Bsig}$ using a projection operator $\hat{P}^{(\Op)}_{\Bsig}$ (defined in the same manner as Eq. (\ref{projection1}) to truncate support of states to $\Hil^{(\Op)}_{\Bsig}$ only), but this time, \emph{without} normalizing the result of the projection, so that we can explicitly measure the support in $\Hil^{(\Op)}_{\Bsig}$. We see that
\begin{multline}
\label{pi_not1_psi}
\hat{P}^{(\Op)}_{\Bsig} \ket{\psi(t)} =  - \frac{i t}{\braket{\psi |\hat{P}^{(\Op)}_{\Asig}|\psi}} \: \tunnel \hat{P}^{(\Op)}_{\Asig} \spsi\\
 - \frac{t^2}{2 \braket{\psi |\hat{P}^{(\Op)}_{\Asig}|\psi}} \:  \left( \ham_{\Bsig}\tunnel + \tunnel \ham_{\Asig} \right) \hat{P}^{(\Op)}_{\Asig} \spsi  \: \in \Hil^{(\Op)}_{\Bsig} \subset \Hil \: .
\end{multline}
The support of $\ket{\psi(t)}$ in $\Hil^{(\Op)}_{\Bsig}$ is given by the overlap of Eq. (\ref{pi_not1_psi}) with the time-evolved state $\ket{\psi(t)}$ itself, which is, to $\mathcal{O}(t^{2})$,
\begin{equation}
\braket{\psi(t)| \hat{P}^{(\Op)}_{\Bsig}| \psi(t) } \: =  \: \frac{\braket{\psi  |  \hat{P}^{(\Op)}_{\Asig} \tunnel^{2} \hat{P}^{(\Op)}_{\Asig} | \psi}}{\braket{\psi|\hat{P}^{(\Op)}_{\Asig}|\psi} ^{2}} \: t^2.
\end{equation}
The coefficient of $t^2$ defines the \emph{Tunneling Spread} $\mathbb{T}\bigl( \spsi, \Asig, D_{\oplus} \bigr)$ of the state $\spsi$ in the subspace $\Hil^{(\Op)}_{\Asig}$ in the decomposition $D_{\oplus}(\Op,\sigma,m)$:
\begin{equation}\label{eq:t_spread}
\mathbb{T}(\spsi, \Asig, D_{\oplus}) \equiv  \frac{1}{2}\frac{d^2}{dt^2} \bigl(\braket{\psi(t)| \hat{P}^{(\Op)}_{\Bsig}| \psi(t) }\bigr) \: = \: \frac{\braket{\psi  |  \hat{P}^{(\Op)}_{\Asig} \tunnel^{2} \hat{P}^{(\Op)}_{\Asig} | \psi}}{\braket{\psi|\hat{P}^{(\Op)}_{\Asig}|\psi} ^{2}}  = \braket{\psi_{\Asig} | \tunnel^{2} | \psi_{\Asig}}.
\end{equation}
The tunneling spread is a time-independent quantity but characterizes the robustness of initially localized states under time evolution in a given direct-sum decomposition. It is evident that the tunneling Hamiltonian plays a crucial role in determining the spread of localized states in the direct-sum. Note also the strong dependence on the choice of state $\spsi$ and decomposition $D_{\oplus}$ (and hence, $\hat{P}^{(\Op)}_{\Asig}$), as expected. 

Before closing this section, we define two other important quantities which will be used in our toy model of Section \ref{sec:double-well} below. As noted in Section \ref{sec:direct-sum}, we label the energy eigenstates of $\ham$ as $\{\ket{e_{i}} \} \: ,\: i = 1,2,\ldots,N $ with corresponding energies $E_{i}$, respectively.
We will be interested in studying the evolution of energy eigenstates projected down to $\Hil^{(\Op)}_{\Asig}$ using a projection operator $\hat{P}^{(\Op)}_{\Asig}$ (as defined in Eq. \ref{projection1}). Define the normalized, projected energy eigenstates by
\begin{equation}
\label{En_proj_state}
\ket{E_{n}}_{\Asig} \equiv \frac{\hat{P}^{(\Op)}_{\Asig} \ket{E_{n}}}{\braket{E_{n} |\hat{P}^{(\Op)}_{\Asig}|E_{n}}}  \in \Hil^{(\Asig)}_{\Asig} \subset \Hil.
\end{equation}
These states have energy expectation values
\begin{equation}
\label{En_proj}
\left( E_{n} \right)_{\Asig} \equiv  \: \:   \frac{\braket{E_{n} |\hat{P}^{(\Op)}_{\Asig} \ham \hat{P}^{(\Op)}_{\Asig} |E_{n}}}{\braket{E_{n} |\hat{P}^{(\Op)}_{\Asig}|E_{n}}^{2}} \: .
\end{equation}

We would also like to quantify the degree of scrambling of a direct-sum decomposition as a whole.
The tunneling spread captures the robustness of an individual state by taking the expectation value of $\ham^{2}_{\mathrm{tunnel}(\sigma)}$ with respect to the projected state.
As a quantifier of how scrambled our decomposition is, we take the trace $\Tr\left(\ham^{2}_{\mathrm{tunnel}(\sigma)} \right)$ in the basis $\{\ket{o_j} \} $ of $\Op$ eigenstates :
\begin{equation}
\label{tunneltrace}
\Tr\left(\ham^{2}_{\mathrm{tunnel}(\sigma)} \right) = \sum_{j = 1}^{N} \braket{o_{j} | \ham^{2}_{\mathrm{tunnel}(\sigma)} | o_{j}} = \sum_{j = 1}^{N} \(\mathbb{T}(\ket{o_{j}}, \Asig, D_{\oplus}) +\mathbb{T}(\ket{o_{j}}, \Bsig, D_{\oplus}) \).
\end{equation}
The last equality follows because the projectors act trivially on the $\{\ket{o_j} \} $.
\section{A Worked Example: The Double-Well Potential}\label{sec:double-well}
In this section, we apply this construction of direct sums and direct-sum locality to a simple, concrete example: the quantum-mechanical double-well potential. While the usual construction of the double-well potential in standard, non-relativistic quantum mechanics textbooks is based on an infinite-dimensional Hilbert space $\mathbb{L}^{2}(\mathbb{R})$ with position and momentum operators, we will construct an analogous finite-dimensional version of the same, in line with our motivation for considering locally finite-dimensional Hilbert spaces of the type relevant for quantum gravity. As we will see, the double-well potential plays very naturally with the direct-sum decomposition and can be used to illustrate features of direct-sum locality very cleanly.

\begin{figure}[t]
  \begin{center}
    \includegraphics[width=1\textwidth]{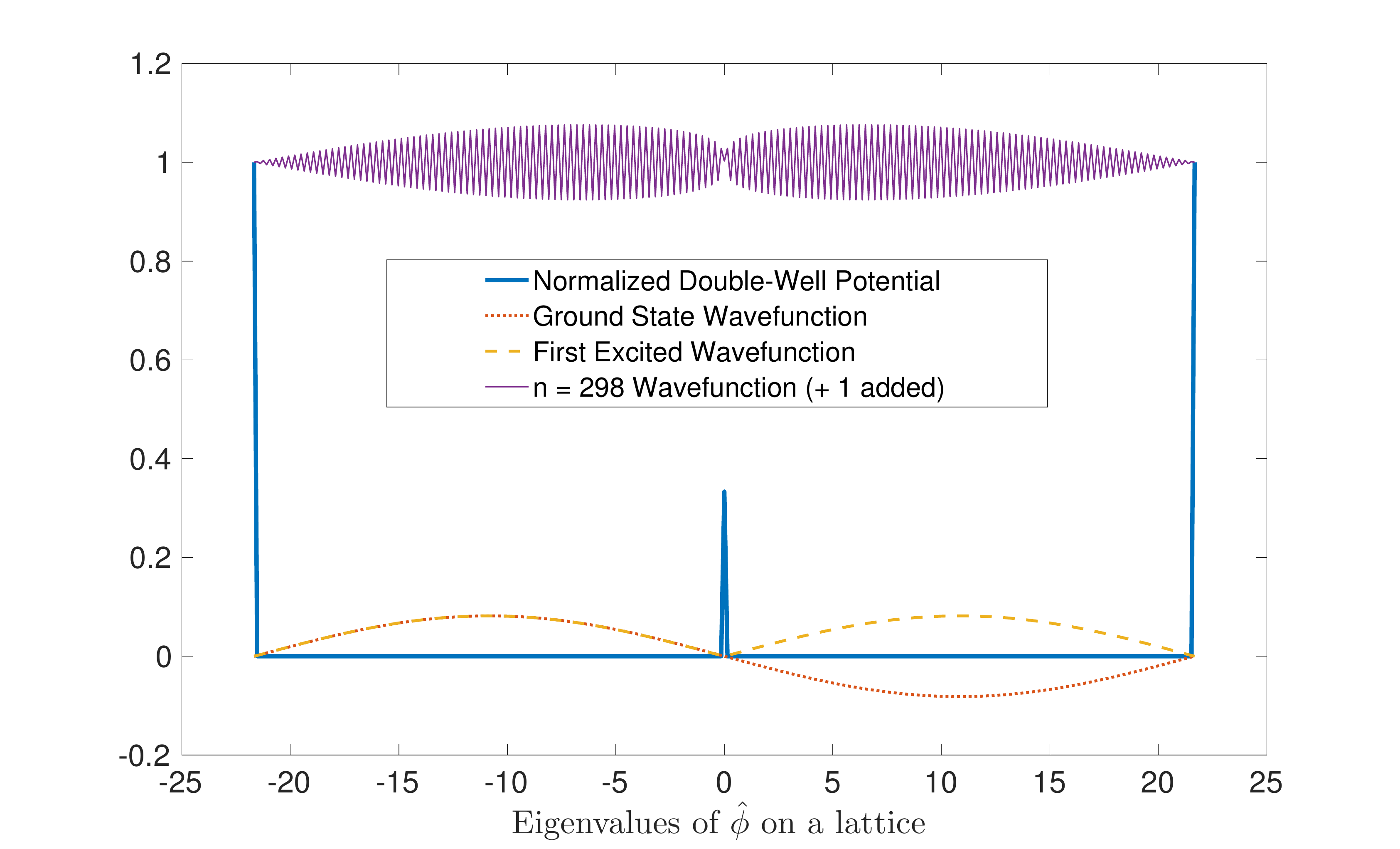}
  \end{center}
  \caption{The $\opphi$-representation of the double-well potential $\hat{V}(\opphi)$ for a Hilbert space of dimension $N = 301$ is plotted. Along with it, we show the lowest two energy eigenstates and one of the highest ones ($n=298$) for a Hilbert space corresponding to $N=301$. We have added $+1$ by hand to the wave function of the $n=298$ state to cleanly separate it from the low-lying one and demonstrate how higher energy states are delocalized across the two wells. Eigenstates $n=299,300, 301$ are very high energy states localized near the barriers of the potential as a result of the the finite-dimensional, cyclic structure of $\opphi$, as can be seen in Fig. \ref{fig:dwspec}. The state $n=299$ is peaked around the central barrier, and $n=300, 301$ are peaked at the edges.}
    \label{fig:dw_potential_states}
\end{figure}

Let us define the Hamiltonian for our double-well system in the standard way,
\begin{equation}
\label{H_dw}
\ham = \frac{\oppi^{2}}{2} + \hat{V}(\opphi),
\end{equation}
where $\oppi$ and $\opphi$ are finite-dimensional analogues of the momentum and position operators which we will define below.
The ``potential" $\hat{V}(\opphi)$ is taken to be,
\begin{equation}
\label{V_dw}
\hat{V}(\opphi) = \begin{cases}
V_{0} & \: \: \: \: \:  \phi_{j} = 0  \\
V_{\mathrm{edge}} & \: \: \: \: \:  \phi_{j} = \pm l \Delta\phi  \\
0 & \: \: \: \: \: \mathrm{elsewhere} \: ,
\end{cases}
\end{equation}
where $V_{0}$ is a positive, real number representing the central barrier potential and $V_{\mathrm{edge}}$ is the positive potential at the edges of our $\phi$-lattice. The separation between eigenvalues of $\opphi$ is denoted by $\Delta \phi$ which is defined in Eqs. (\ref{phi_delta_eig}) and (\ref{CCRconstraint}) below. We have numerically implemented this system in MATLAB for a Hilbert space of dimension $N = 301$, with $V_{\mathrm{edge}} = 3V_{0} = 10 || \oppi^{2} / 2 ||_{2}$ to ensure that the central barrier is lower than the edge barriers. In Figure \ref{fig:dw_potential_states}, we plot this double-well potential and show the the lowest two and one of the higher energy eigenstates represented in $\phi$-space. As expected and well-known from quantum mechanics, the low-lying states represent superpositions of localized states within each well, whereas higher energy states are delocalized over the full double well. 

\begin{figure}[t]
  \begin{center}
    \includegraphics[width=1\textwidth]{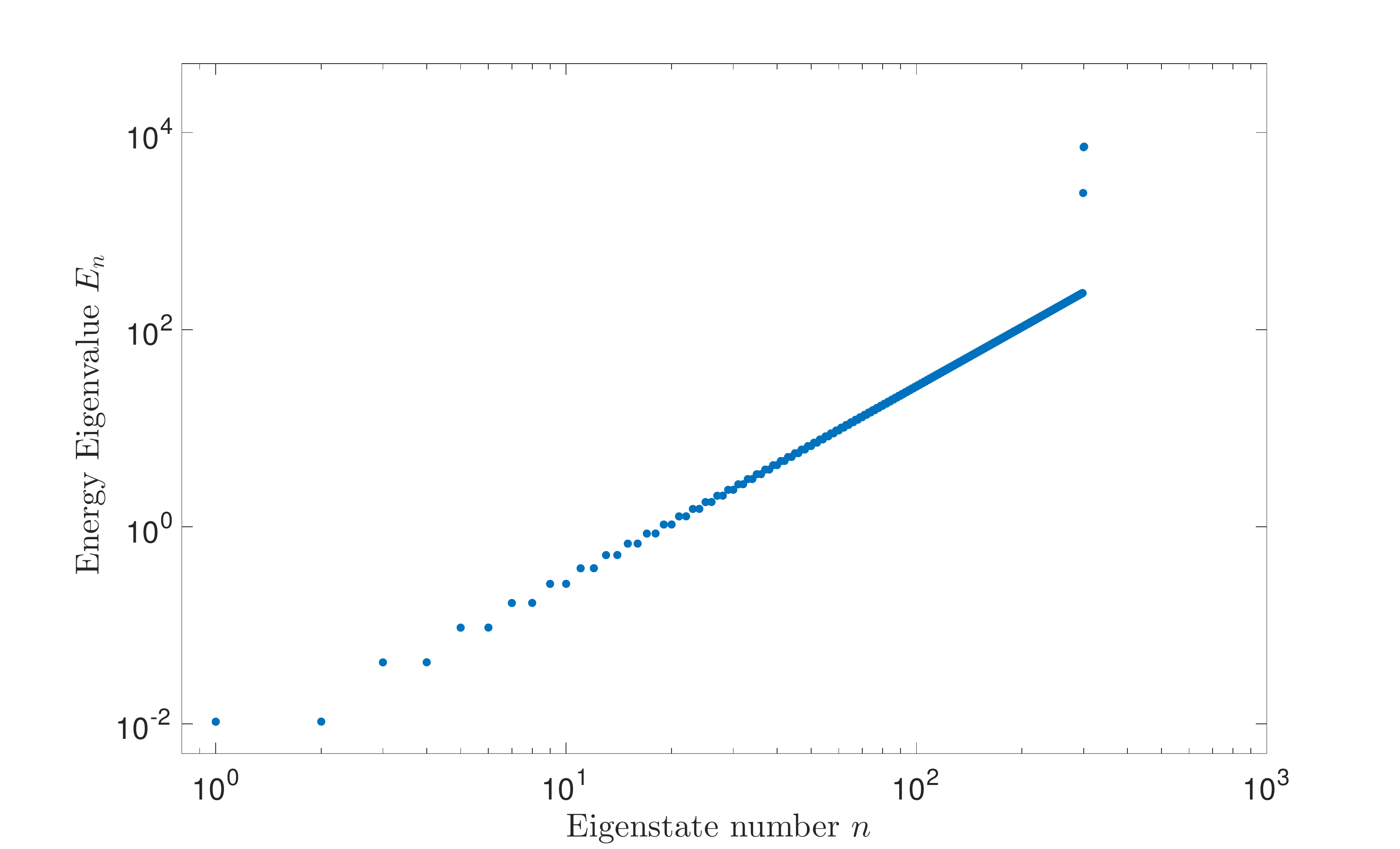}
  \end{center}
  \caption{Energy spectrum of the double-well Hamiltonian of Eq. (\ref{H_dw}). One can easily notice the (approximate) double degeneracy in the low-lying eigenvalues (corresponding to the two symmetric wells). The last three (the highest two are indistinguishable on the plot) highly-energetic eigenstates are an artifact of working in a finite-dimensional space with a relatively small $N$ with the algebra of conjugate variables having a cyclic structure. They cause spurious features in other plots but have no bearing on our physical results.}
    \label{fig:dwspec}
\end{figure}

In addition, we plot the energy eigenvalues (spectrum) of the Hamiltonian in Figure \ref{fig:dwspec}, where the expected double degeneracy of the lower eigenvalues is demonstrated. A few of the higher most eigenvalues are exceptionally large; this is a consequence of working with a Hilbert space of a relatively small size ($N=301$ in our case) with a cyclic structure. Such states of extremely high energy will not be explicitly studied here, but they will induce stray effects in the results and plots to follow which do not bear any physical significance on our main results. 

There is an obstacle to defining $\oppi$ and $\opphi$ in the same way as in standard one-dimensional quantum mechanics. It is well known that Heisenberg's canonical commutation relation (CCR) between pairs of conjugate variables has no finite-dimensional irreducible representations. However, its exponential form, given by Weyl \cite{weyl1950theory}, does admit finite-dimensional representations, based on the Generalized Clifford Algebra (GCA) (\cite{Jagannathan:2010sb} for a review), and can be used to construct finite-dimensional conjugate variables which reduce to the usual ones obeying Heisenberg's CCR in the infinite-dimensional limit (see, for example, \cite{Schwinger570,Jagannathan:1981ri,SanthanamTekumalla1976}). We give a short introduction to the GCA as an appropriate construction of finite-dimensional conjugate variables in the Appendix.

The appropriate way to define $\oppi$ and $\opphi$ is therefore by using the GCA. We denote the Hilbert space of our Double-Well (DW) system as $\Hil_{DW}$, with an odd, finite dimension $N = 2l+1$. (The odd dimension is chosen so that  our field variable $\phi$ can lie on a one-dimensional lattice centered around $0$.) On $\mathcal{L}(\Hil_{DW})$, we associate a pair of conjugate variables $\opphi$ and $\oppi$ which form a GCA. From the point of view of the GCA, $\opphi$ and $\oppi$ are on the same footing, so we need to make a choice of which operator to assign to position and which to momentum. As already noted, the operator corresponding to a ``lattice" variable is chosen to be $\opphi$, which has eigenvalues
\be
\label{phi_delta_eig}
\{ \phi_{j} = j\Delta\phi \: \: , j =  -l ,  (-l+1) ,\ldots,0,\ldots,(l-1),l \},
\ee 
where $\Delta\phi$ is a positive real number constrained by the algebra to obey
\be
\label{CCRconstraint}
(2l+1) \Delta\phi \Delta\pi = 2\pi
\ee
and $\Delta\pi$ is the uniform difference between eigenvalues of $\oppi$. 
This constraint ensures that Heisenberg's canonical commutation relation is recovered in the infinite-dimensional $N \to \infty$ limit. In our numerical implementation, we have taken $\Delta\phi = \Delta\pi = \sqrt{2\pi/(2l+1)}$ and $l=150$.

Thus eigenstates of $\opphi$ can be thought of labeling sites on a 1-D lattice with cyclic boundary conditions as specified by the GCA. The conjugate variable to $\opphi$ is $\oppi$ which generates translations in the eigenspace of $\opphi$ (and vice versa). For our purposes, we will use $\opphi$ and $\oppi$ in analogy to position and momentum operators in standard textbook quantum mechanics in one spatial dimension on $\mathbb{L}^{2}(\mathbb{R})$, but here representing bounded operators on a finite-dimensional Hilbert space.

\begin{figure}[t]
  \begin{center}
    \includegraphics[width=1\textwidth]{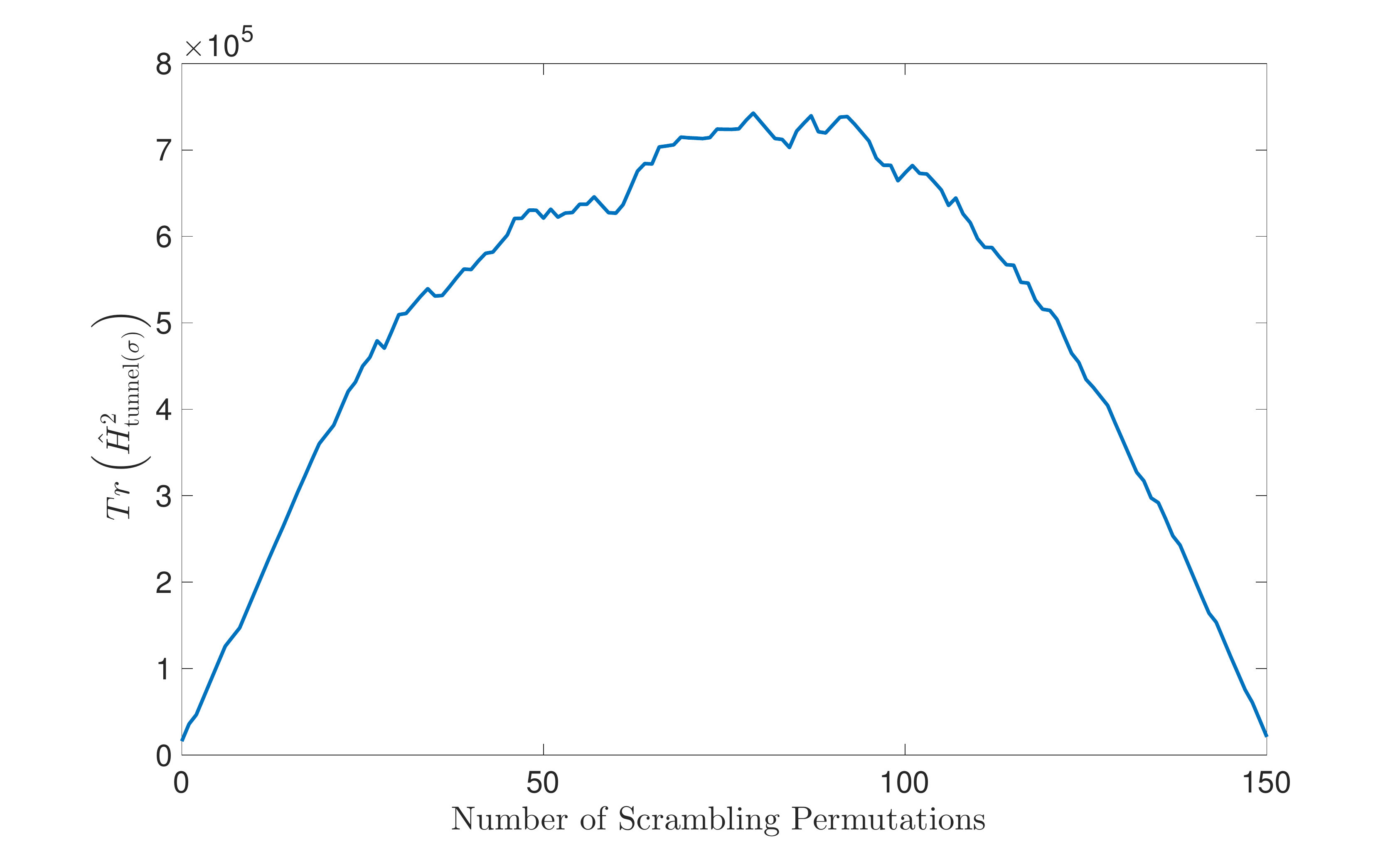}
  \end{center}
  \caption{The trace of the square of the tunneling term in the Hamiltonian $\Tr\left(\ham^{2}_{\mathrm{tunnel}(\sigma)} \right)$ for different decompositions building up from the canonical, ordered $(-l,-l+1,\ldots,-1) \cup  (0, 1,2,\ldots,l)$ decomposition. This quantity acts a measure of the strength of the tunneling term for the choice $\Op \equiv \opphi$ and quantifies the scrambling of the decomposition as discussed in the text.}
    \label{fig:nperm_tunneltrace}
\end{figure}

Having defined the system, we can proceed to study different choices of scramblings $\sigma$. Because the system defines a symmetric double-well potential, we expect that the only good choices are those which keep the size $m$ of $\Asig$ fixed at $m=l$. In particular, we start from the ordered, canonical $\phi$-lattice $\left(-l,-l+1,\ldots,0,\ldots,l-1,l\right)$ and sequentially build up different scramblings by swapping a pair of randomly chosen sites from $\Asig$ and $\Bsig$ separated by the barrier fixed at $\phi = 0$. In the canonical, ordered decomposition, the Hamiltonian is the sum of a local kinetic term and a potential (a highly non-generic feature), and as one scrambles away from it and becomes more non-local, the tunneling term becomes more important. This behavior is seen in Figure \ref{fig:nperm_tunneltrace}, where $\Tr\left(\ham^{2}_{\mathrm{tunnel}(\sigma)} \right)$ (which we defined in Eq. \ref{tunneltrace} in Section \ref{sec:locality} as a measure of how scrambled the decomposition is) correlates with the number of scrambling swaps applied to the canonical, ordered lattice. 

As expected, since we are working with a Hilbert space corresponding to $l = 150$ and swapping pairs of sites sequentially across the central barrier,the decomposition becomes more non-local (and hence has higher $\Tr\left(\ham^{2}_{\mathrm{tunnel}(\sigma)} \right)$) until reaching $\approx l/2$ swaps, after which we start approaching the case where the two wells are swapped entirely (up to internal scramblings within each well which are inconsequential for our purposes, since the potential energy defined on such well configurations is zero and the Hamiltonian reduces to just a local kinetic term within each well).

\begin{figure}[t]
  \begin{center}
    \includegraphics[width=1\textwidth]{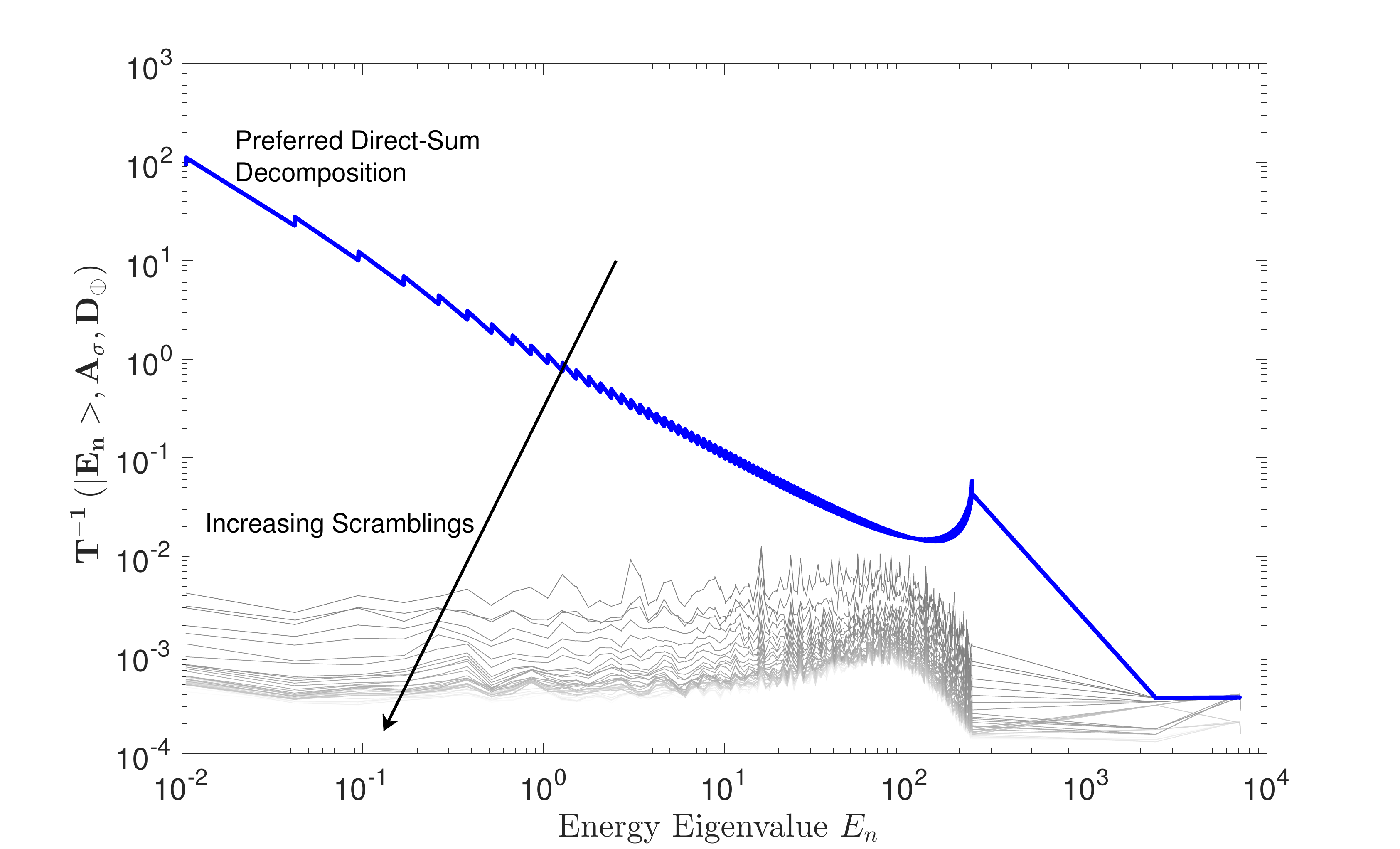}
  \end{center}
  \caption{Plot showing the dependence of Tunneling Spread $\mathbb{T}\bigl( \spsi, \Asig, D_{\oplus} \bigr)$ on energy eigenstates (represented by their eigenvalues $E_{n}$) of the double-well Hamiltonian expressed in different direct-sum decompositions $D_{\oplus}(\opphi,\sigma,m=l)$ of $\Hil_{DW}$. There exists a preferred decomposition into direct sum subspaces (the canonical, ordered $(-l,-l+1,\ldots,-1) \cup  (0, 1,2,\ldots,l)$ one) in which low-lying energy states have very small tunneling spreads and hence represent robust, localized states in the direct-sum subspace. Other decompositions are near-generic where there is no manifestation of direct-sum locality. Notice the log scale on the y-axis representing the tunneling spread. The very-high energy behavior is a consequence of the three largest energy eigenstates, which are artifacts of finite-dimensional, cyclic constructions and do not bear any physical significance for our results.}
    \label{fig:tspread_En}
\end{figure}

We next study the properties of individual energy eigenstates $\ket{E_{n}}$ of the double-well Hamiltonian of Eq. (\ref{H_dw}) and their projected counterparts $\ket{E_{n}}_{\Asig}$ on $\Hil^{(\Op)}_{\Asig}$.
We compute the tunneling spread for each projected eigenstates in different choices of direct-sum scramblings, based on the operator $\opphi$ but varying $\sigma$ by swapping sites in the same way as described above. The results are plotted in Figure \ref{fig:tspread_En}. 

We see that there exists a preferred decomposition of $\Hil_{DW}$ into two subspaces based on $\opphi$, in which the low-lying energy states offer a very natural set of robust, localized states within the $\Hil^{(\opphi)}_{\Asig}$ direct-sum subspace, acting as semiclassical states which maintain their support under evolution by the Hamiltonian. Changing the scrambling even slightly destroys the strong correlation we see for the preferred decomposition. As one goes on further scrambling the canonical decomposition, by continually swapping lattice sites across the central barrier, the tunneling spread systematically increases, but occupies a ``degenerate band" indicating how generically non-local arbitrary decompositions are. In addition, as expected, even with the correct choice of decomposition it is only the low-lying states which exhibit robustness and locality. Higher-energy states are already delocalized to begin with (as seen in Figure \ref{fig:dw_potential_states}) and approach the generic band of states for any choice of scramblings. 

\begin{figure}[t]
  \begin{center}
    \includegraphics[width=1\textwidth]{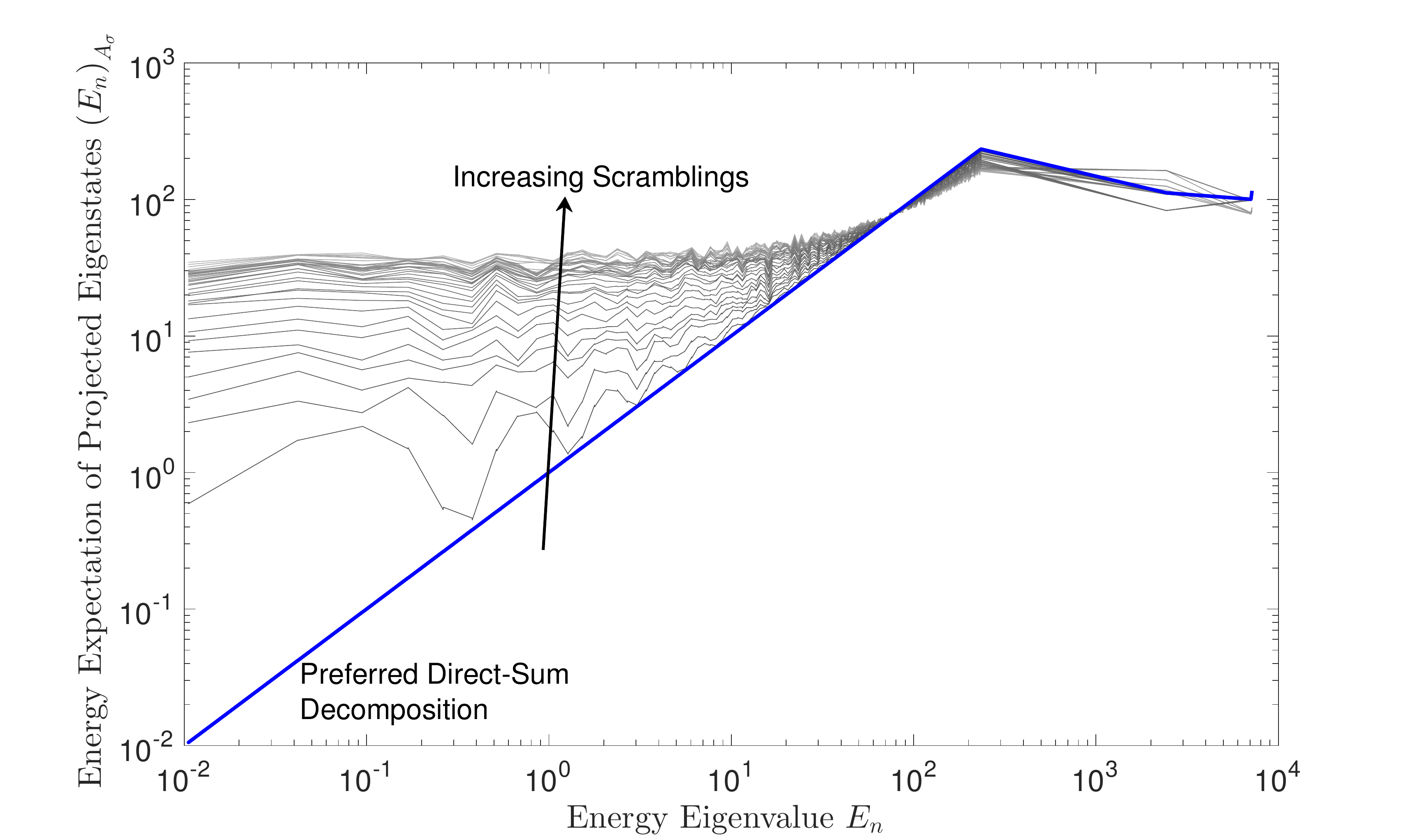}
  \end{center}
  \caption{Plot showing correlation between $E_{n}$ and $\left(E_{n}\right)_{\Asig}$ for different choices of scramblings $D_{\oplus} = \{\opphi,\sigma,m\}$ in the double-well toy model. Only in the canonical, preferred decomposition do the low-lying eigenstates of $\ham$ serve as low energy states which can be localized within a given direct-sum subspace. The very-high energy behavior is a consequence of the three largest energy eigenstates, which are artifacts of finite-dimensional, cyclic constructions and do not bear any physical significance for our results.} 
    \label{fig:En_Enproj}
\end{figure}

Such ideas can be further understood by studying the correlation between $E_{n}$, the energy eigenvalues of the Hamiltonian, and the expectation value of the energy of the \emph{projected} states, $\ket{E_{n}}_{\Asig}$, defined as $\left(E_{n}\right)_{\Asig}$ in Eq. (\ref{En_proj}) above. As shown in Figure \ref{fig:En_Enproj}, in the canonical, preferred decomposition discussed above, low-lying eigenstates of the Hamiltonian inherit their low energy features when projected down to the $\Hil^{(\opphi)}_{\Asig}$ subspace and can describe localized states which are robust under time evolution (as described by by Figure \ref{fig:tspread_En}). For decompositions successively more scrambled from the canonical, the low-lying eigenstates lose their role as localized states because they are mapped to energetically unfavoured states once projected onto the direct-sum subspace. This further demonstrates how arbitrary decompositions place all states on equal footing once projected; only in the preferred decomposition, which respects the system's dynamics, are localization features in the low-lying states made manifest.

Now that we have discussed ideas of direct-sum locality in a concrete example of the double-well potential and seen the interesting, non-generic features of the preferred direct-sum decomposition, let us interpret these ideas in the context of cosmology and field theory.
\section{Towards Field Theory}\label{sec:field-theory}

The toy model considered in the previous section demonstrated features that we expect are generic in situations where a lattice theory on a semiclassical spatial background can be described on direct-sum sectors of the Hilbert space. In particular, they illustrate how to select a preferred direct-sum decomposition under which the low-lying states of the Hamiltonian can represent robust, semiclassical states which remain localized under time evolution.

It is important to point out, however that in our construction above there was no notion of tensor factorization of Hilbert space. 
Explicitly, the analog of position in our finite-dimensional toy model was labeled by the eigenvalues of the operator $\hat \phi$.
As we take the dimension of the Hilbert space to infinity to recover one-dimensional quantum mechanics, $\hat \phi$ reduces to the position operator $\hat x$. 
There is thus no real sense in which degrees of freedom in the Hilbert space of the model are actually localized at a given position.
Contrast this with field-theory, in which the analogous operators measure the field value at a particular point, $\hat \phi_x$. 

In other words, our toy model was a first quantized, not a second quantized theory, and we should not expect either the left or right Hilbert subspaces to factorize in the manner required of a lattice field theory.
Its goal was to demonstrate, as a proof of principle, how there can exist a preferred direct-sum decomposition for a given Hamiltonian for which we can recover robust, semi-classical structure in different direct-sum subspaces. 
To extend these ideas to Hilbert spaces where lattice structure exists, we first require an appropriate number-theoretic division into direct sums in which each sum allows a tensor factorization, as discussed in Section \ref{sec:non-genericity} above. 
However, even once we have checked that such a tensor factorization can exists, the success of the decoherence program within each such direct-sum factor requires a specific, non-generic direct-sum decomposition which is chosen based on locality characteristics governed by the Hamiltonian.
It is this choice of decomposition which our results on the toy model in the previous section can inform.

In Section \ref{sec:direct-sum} above, we suggested that theories in which each term in the direct sum describes a geometric field theory occur frequently in quantum field theory in curved space.
Whenever we have a theory that describes multiple (meta)stable vacua, such those invoked in landscape eternal inflation \cite{Garriga:1997ef}, string-theoretic mechanisms for producing de~Sitter solutions \cite{Kachru:2003aw}, or for anthropic purposes \cite{Bousso:2000xa}, we should decompose the Hilbert space of the theory into a sum of subspaces which describe field theory around the geometry corresponding to each individual vacuum in the theory (see \cite{Nomura:2011dt,Nomura:2011rb} for related ideas).
Interactions between the subspaces, which we have referred to as tunneling terms, play the dual role of describing transitions between the vacua and determining which states of the field theories remain robust under time evolution.
They also determine what the ultimate ground state of the theory is, as states initially localized within a particular vacuum eventually relax to a particular superposition of states across all of the vacua \cite{Boddy:2014eba}.

Let's finally return to the questions we asked in the Introduction. Given only the Hilbert-space data of a theory--that is, its dimension and the spectrum of its Hamiltonian--how could we deduce that such a theory describes a landscape of field theories of this sort?
And how could we identify the local degrees of freedom within each field theory?
The answer we have presented in the last several sections is to vary across the operators $\Op$ and choice of decompositions $(\sigma,m)$ which define a direct-sum decomposition $D_{\oplus}(\Op,\sigma,m)$. 
(If we are considering more arbitrary decompositions into more than two pieces, we should replace the size $m$ with the dimensionalities $\{m_i\}$ of each subspace.)
Then we should compare the measures of direct-sum locality we defined in Section \ref{sec:locality} above---in particular, the tunneling spread $\mathbb{T}\bigl( \spsi, \Asig, D_{\oplus} \bigr)$ (\ref{eq:t_spread}) and the size of the tunneling term $\Tr\left(\ham^{2}_{\mathrm{tunnel}(\sigma)} \right)$ (\ref{tunneltrace})---across different decompositions.
Once we have identified a particular decomposition in which low-lying states in a particular subspace remain robust under the action of the Hamiltonian, such as the canonical decomposition into left and right in our toy model above, we can use the familiar methods of decoherence within the subspaces to identify the local degrees of freedom.
That is, we can write each subspace as a tensor product of smaller factors and identify the choice of basis in which interactions between the factor act like a monitoring environment.

\section{Conclusion}\label{sec:conclusion}

In this paper we have tried to take some preliminary steps towards understanding the conditions under which lattice or geometrical structures emerge from Hilbert-space dynamics.
We have introduced the notion of \emph{direct-sum decompositions} of a Hilbert space which partition the Hilbert space into multiple pieces spanned by subsets of the eigenstates of an operator.
We argued that preferred decompositions which allow the possibility of local degrees of freedom can be identified by the existence in such decompositions of states with low \emph{tunneling spread}, which remain robust under time evolution.
We studied the selection of a canonical decomposition in a simple toy model, the finite-dimensional discretization of the double-well potential.

Much work remains to be done in order to fill out the entire research program we mentioned in the Introduction: understanding how the classical, geometric world we observe emerges from a fundamental Hilbert space. 
With respect to our particular problem, it would be interesting to more explicitly understand under what conditions landscapes of vacua, which seem to be ubiquitous in our models of quantum gravity, can emerge. 
In particular nothing in the direct-sum understanding we have sketched in this paper seems to preclude the different metastable vacua from differing dramatically, for example in having different numbers of fundamental fields or even differing numbers of dimensions.
In addition, there is some tension between a description of spacetime with a lattice structure and the existence of gauge symmetries or diffeomorphism invariance \cite{donnelly+giddings2016_1,donnelly+giddings2016_2,Donnelly:2016auv,Speranza:2017gxd,Witten:2017hdv}, which might be solved by introducing additional ``edge modes'' in addition to degrees of freedom located at lattice sites. 
Describing the Hilbert spaces of such theories as direct-sum decompositions would require additional generalization.

\section*{Acknowledgments}
We are thankful to ChunJun (Charles) Cao, Sean Carroll, Tamir Hemo and Ingmar Saberi for helpful discussions. J.P. is supported in part by the Simons Foundation and in part by the Natural Sciences and Engineering Research Council of Canada. This material is based upon work supported by the U.S. Department of Energy, Office of Science, Office of High Energy Physics, under Award Number DE-SC0011632, as well as by the Walter Burke Institute for Theoretical Physics at Caltech and the Foundational Questions Institute. 

\appendix
\label{sec:appendix}
\section{A Primer on Generalized Clifford Algebra}
 Consider a finite-dimensional Hilbert Space $\hs$ of dimension $\Dim \hs = N \in \mathbb{Z}^{+}$ with $N < \infty$. A Generalized Clifford Algebra(GCA) on the space of linear operators $\mathcal{L}(\hs)$ acting on $\hs$ comes equipped with two unitary (but not necessarily hermitian) operators as generators of the algebra, call them $\hat{A}$ and $\hat{B}$ which satisty the following commutation relation,
\begin{equation}
\label{weylbraid}
\hat{A}\hat{B} = \omega\hat{B}\hat{A} \: ,
\end{equation}
where $\omega = \exp\left(2 \pi i /N\right)$ is the $N$-th primitive root of unity. This commutation relation is also more commonly known as the Weyl Braiding relation and any further notions of commutations between conjugate, self-adjoint operators (which will be defined from $\hat{A}$ and $\hat{B}$) will be derived from this fundamental braiding relation. In addition to being unitary, $\hat{A}\hat{A}^{\dag} = \hat{A}^{\dag}\hat{A} =  \eye  = \hat{B}\hat{B}^{\dag} = \hat{B}^{\dag}\hat{B}$, the algebra cyclically closes, giving it a cyclic structure in eigenspace,
\begin{equation}
\label{AdBdI}
\hat{A}^{N} = \hat{B}^N = \eye \: ,
\end{equation} 
where $\eye$ is the identity operator on $\mathcal{L}(\hs)$. 
The GCA can be constructed for both even and odd values of  $N$ and both cases are important and useful in different contexts. Here, we focus on the case of odd $N \equiv 2 l + 1$ which will be useful in constructing conjugate variables whose eigenvalues can be thought of labelling lattice sites, centered around $0$. While all of the subsequent construction can be done in a basis-independent way, we choose a hybrid route, routinely switching between an explicit representation of the GCA and abstract vector space relations. Let us follow the convention that all indices used in this section(for the case of odd $N = 2l+ 1$), for labelling states or matrix elements of an operator in some basis etc. will run from $-l ,(-l + 1) ,\ldots,-1,0,1,\ldots,l$. The operators are further specified by their eigenvalue spectrum, and it is identical for both the GCA generators $\hat{A}$ and $\hat{B}$,
\begin{equation}
\label{specABodd}
\mathrm{spec}(\A) \: = \: \mathrm{spec}(\B) \: = \: \{\omega^{-l}, \omega^{-l + 1},\ldots, \omega^{-1}, 1, \omega^{1}, \ldots, \omega^{l-1}, \omega^{l}  \} \: .
\end{equation}
There exists a unique irreducible representation (up to unitary equivalences) (see review \cite{Jagannathan:2010sb} for details) of the generators of the GCA defined via Eqs. (\ref{weylbraid}) and (\ref{AdBdI}) in terms of $N \times N$ matrices
\begin{equation}
\label{Amatrix}
  A \: = \:      \begin{bmatrix}
       0  & 0  & 0 & \ldots  & 1          \\[0.3em]
        1  & 0  & 0 &  \ldots   & 0          \\[0.3em]
        0  & 1  & 0 &  \ldots   & 0          \\[0.3em]
       . & .  & \ldots   & .          \\[0.3em]
              .  & .  & \ldots   & .          \\[0.3em]
        0  & 0   & \ldots   & 1 & 0          \\[0.3em]
     \end{bmatrix}_{N \times N} \: .
\end{equation}
\begin{equation}
\label{Bmatrix}
  B \: = \:      \begin{bmatrix}
       \omega^{-l}  & 0  & 0 & \ldots  & 0          \\[0.3em]
        0  & \omega^{-l+1}  & 0 &  \ldots   & 0          \\[0.3em]
       . & .  & \ldots   & .          \\[0.3em]
              .  & .  & \ldots   & .          \\[0.3em]
        0  & 0  & 0 & \ldots   & \omega^{l}          \\[0.3em]
     \end{bmatrix}_{N \times N} \: .
\end{equation}
The $\hat{.}$ has been removed to stress that these matrices are representations of the operators $\A$ and $\B$ in a particular basis, in this case, the eigenbasis $\B$(so that $B$ is diagonal). More compactly, the matrix elements of operators $\A$ and $\B$ in the basis representation of eigenstates of $\B$,
\begin{equation}
\left[ A \right]_{jk} \equiv \braket{b_{j} | \A | b_{k}} =  \delta_{j,k+1} \: ,
\end{equation}
\begin{equation}
\left[ B \right]_{jk} \equiv \braket{b_{j} | \B | b_{k}}  = \omega^{jk} \delta_{j,k} \: ,
\end{equation}
with the indices $j$ and $k$ running from $-l, \ldots , 0, \ldots, l$ and $\delta_{jk}$ is the Kronecker Delta function. 
Consider the set $\{\ket{b_j}\}, \: j = -l,\ldots,l$ of states to be the set of eigenstates of $\hat{B}$, 
\begin{equation}
\label{Baction}
\hat{B} \ket{b_j} = \omega^j \ket{b_j} \: , \: j = -l,\ldots,0,\ldots,l
\end{equation}
As can be evidently seen in the matrix representation of $\A$ in Eq. (\ref{Amatrix}), the operator $\hat{A}$ acts as a a ``cyclic shift" operator for the eigenstates of $\B$, sending an eigenstate to the next,
\begin{equation}
\label{Aaction}
\hat{A}\ket{b_j} = \ket{b_{j+1}} \: .
\end{equation}
The unitary nature of these generators implies a cyclic structure which identifies $\ket{b_{l+1}} \equiv \ket{b_{-l}}$, so that $\hat{A}\ket{b_l} = \ket{b_{-l}}$. The operators $\A$ and $\B$ have the same relative action on one another's eigenstates, since nothing in the algebra sets the two apart. It has already been seen in Eq. (\ref{Aaction}) that $\A$ generates (unitary, cyclic) unit shifts in eigenstates of $B$ and the opposite holds too: the operator $\B$ generates unit shifts in eigenstates of $\A$ (given by the relation $\hat{A} \ket{a_k} = \omega^k \ket{a_k} \: , \: k = -l , \ldots 0, \ldots l$) and has a similar action with a cylic correspondence to ensure unitarity,
\begin{equation}
\label{BactionOnAstates}
\hat{B}\ket{a_k} = \ket{a_{k+1}} \: ,
\end{equation}
with cyclic identification $\ket{a_{l+1}} \equiv \ket{a_{-l}}$. Hence we already have a set of operators which generate shifts in the eigenstates of the other, which is precisely what conjugate variable do and which is why we see that the GCA provides a very natural structure to define conjugate variables on Hilbert Space.
The GCA generators $\A$ and $\B$ have been extensively studied in various contexts in quantum mechanics, and are often referred to as  ``Clock and Shift" matrices in the literature and offer a higher dimensional, non-hermitian generalisation of the Pauli matrices. In particular, for sake of completeness, we mention that for $N = 2$, it will be seen that $A = \sigma_{1}$ and $B = \sigma_{3}$ which recovers the famous Pauli matrices. 
\\
\\
The defining notion for a pair of conjugate variables is the identification of two self-adjoint operators acting on Hilbert space, each of which generates translations in the eigenstates of the other. For instance, in (conventionally infinite-dimensional) textbook quantum mechanics, the momentum operator $\hat{p}$ generates shifts/translations in the eigenstates of its conjugate variable, the position $\hat{q}$ operator and vice versa. Taking this as our defining criterion, we would like to define a pair of conjugate operators $\opphi$ and $\oppi$, acting on a finite-dimensional Hilbert space, each of which is the generator of translations in the eigenstates of its conjugate, with the following identification,
\begin{equation}
\label{phipidef}
\A \equiv \exp{(-i \alpha \oppi)} \: , \: \: \: \: \B = \exp{(i \beta \opphi)} \: ,
\end{equation}
where $\alpha$ and $\beta$ are, non-zero real parameters which set the scale of the eigen-spectrum of the operators $\opphi$ and $\oppi$. They are bounded operators on $\mathcal{\hs}$ and due to the virtue of the GCA generators $\A$ and $\B$ being unitary, the conjugate operators $\opphi$ and $\oppi$ are self-adjoint satisfying $\opphi^{\dag} = \opphi$ and $\oppi^{\dag} = \oppi$. The operator $\oppi$ is the generator of translations of $\opphi$ and vice versa. Of course, $\opphi$ has common eigenstates with those of $\B$ and $\oppi$ shares eigenstates with $\A$. Let us, for the sake of clarity and convenience, label the eigenstates of $\opphi$ as $\ket{\phi_{j}}$ and those of $\oppi$ as $\ket{\pi_{j}}$ with the index $j$ running from $-l,\ldots,0,\ldots,l$. The corresponding eigenvalue equations for $\opphi$ and $\oppi$ can be easily deduced using Eqs. (\ref{phipidef}) and (\ref{specABodd}),
\begin{equation}
\label{phi_eig}
\opphi \ket{\phi_j} = j \left( \frac{ 2 \pi}{(2 l + 1) \alpha} \right) \ket{\phi_j} \: , \: \: \: j = -l, \ldots , 0,\ldots,l \: ,
\end{equation}
\begin{equation}
\label{pi_eig}
\oppi \ket{\pi_j} = j \left( \frac{ 2 \pi}{(2 l + 1) \beta} \right) \ket{\pi_j} \: , \: \: \: j = -l, \ldots , 0,\ldots,l \: ,
\end{equation}
These conjugate variables defined on a finite-dimensional Hilbert space will \emph{not} satisfy Heisenberg's Canonical Commutation relation (CCR) $ \left[ \opphi,\oppi \right] = i $ (in units where $\hbar = 1$) since as is well known by the Stone-von Neumann theorem, there are no finite-dimensional representations of Heisenberg's CCR. However, $\opphi$ and $\oppi$ still serve as a robust notion of conjugate variables and their commutation can be derived from the more fundamental Weyl Braiding Relation of Eq. (\ref{weylbraid}). In the large dimension limit $N \to \infty$, one recovers Heisenberg's form of the CCR if the parameters $\alpha$ and $\beta$ are constrained to obey $\alpha \beta = 2\pi/N$. This completes our lightning review of GCA and conjugate variables on a finite-dimensional Hilbert space.
\bibliographystyle{utphys}
\bibliography{directsumrefs}

\end{document}